\documentclass[english]{article}
\usepackage[latin9]{inputenc}
\usepackage{color}
\usepackage{textcomp}
\usepackage{amstext}
\usepackage{amssymb}
\usepackage{graphicx}
\usepackage{esint}

\makeatletter

\newcommand{\lyxmathsym}[1]{\ifmmode\begingroup\def\b@ld{bold}
  \text{\ifx\math@version\b@ld\bfseries\fi#1}\endgroup\else#1\fi}

\newcommand{\lyxaddress}[1]{
\par {\raggedright #1
\vspace{1.4em}
\noindent\par}
}

\makeatother

\usepackage{babel}
\begin{document}

\title{\textbf{Tricritical quantum point and inflationary cosmology}}

\author{\textbf{Lawrence B. Crowell}}

\maketitle

\lyxaddress{$ $Alpha Institute of Advanced Study 10600 Cibola Lp 311 NW Albuquerque,
NM 87114 also 11 Rutafa Street, H-1165 Budapest, Hungary, email: \textcolor{blue}{lcrowell@swcp.com}; }
\begin{abstract}
The holographic protection due to inflationary cosmology is a consequence
of a quantum tricritical point. In this scenario a closed spacetime
solution transitions into an inflationary de Sitter spacetime. Saturation
of the holographic entropy bound is prevented by the phase change
in the topology of the early universe.

\textbackslash{}footnote\{This essay received an \textbackslash{}honorable
mention\char`\"{} in the 2012 Essay Competition of the Gravity Research
Foundation.\} \textbackslash{}vfill\textbackslash{}eject
\end{abstract}

\section{Introduction}

The role of entropy bounds, holography and the accelerated universe
constitutes a considerable body of work. Fischler and Susskind proposed
ways the holographic principle (HP) does not conflict with inflation
and the accelerated cosmology {[}1{]}. The holographic principle is
satisfied for entropy/energy ratios for bounded systems with event
horizons {[}2{]}. The closed Friedman-Lemaitre-Robertson-Walker (FLRW)
spacetime violates the HP. A proposed cosmological model in {[}3{]}
advances how the HP may accommodate inflation and dark energy, and
derives the equation of state for vacuum energy and pressure {[}4{]}.
It is thought the HP is more fundamental than inflation, where the
HP constructs states on a surface one dimension smaller than the general
spacetime, while inflation is spacetime dynamics based entirely on
some initial conditions. A closed FLRW spacetime will saturate the
entropy bound or HP once the universe reaches its turn around point
to a big crunch{[}5{]}.

The FLRW metric for a closed homogeneous and isotropic cosmology is
\[
ds^{2}=dt^{2}-a^{2}(t)(d\chi^{2}+sin^{2}\chi d\Omega^{2})
\]
for $\chi$ the azimuthal angle of the sphere $\mathbb{S}^{3}$ and
$\Omega$ the steradian measure on any 2-sphere for a given $\chi$.
The particle horizon is 
\[
\chi_{h}~=~\int_{t_{p}}^{t}\frac{dt^{\prime}}{a(t^{\prime})}.
\]
for $t_{p}$ a fiducial start time. The entropy density $\sigma~=~(\rho+p)/T$
is constant and the entropy to area ratio is
\[
\frac{S}{A}~=~\sigma\frac{2\chi_{h}~-~sin(2\chi_{h})}{4a^{2}(\chi_{h})sin^{2}(\chi_{h})}
\]
This saturates for $\chi_{h}=\pi/2$ at maximal expansion of the closed
FLRW cosmology {[}5{]}.

The inflationary model with a scalar inflaton field $\phi$ and potential
$V(\phi)$ evolves as $\dot{\rho}=3(\rho~+~p)=3(1+w)\rho$, which
gives the equation of state parameter $w=-1$ for $\dot{\rho}=0$.
The inflaton field $\phi$ evolves as 
\[
\nabla^{2}\phi~-~\dot{\phi}~-~3H\dot{\phi}~-~\frac{\partial V(\phi)}{\partial\phi}~=~0,
\]
for the Hubble parameter $H~=~\dot{a}/a$. Isotropy $\nabla^{2}\phi~\simeq~0$
simplifies this to $\dot{\phi}~+~3H\dot{\phi}~+~\frac{\partial V(\phi)}{\partial\phi}~=~0$.
The scale factor obeys 
\[
\Big(\frac{\dot{a}}{a}\Big)^{2}~=~H^{2}~=~H_{0}^{2}\Big[\Omega_{\phi}\Big(\frac{a_{0}}{a}\Big)^{3(1+w)}~+~\Omega_{m}\Big(\frac{a_{0}}{a}\Big)^{4}~-~\Omega_{k}\Big(\frac{a_{0}}{a}\Big)^{2}\Big],
\]
where $a_{0}$ and $H_{0}$ are the initial scale factor and Hubble
parameter. $\Omega_{\phi}$, $\Omega_{m}$ and $\Omega_{k}$ are the
density ratios for the inflaton field, relativistic matter or radiation
and the spatial curvature for $k=1$. The equation of state parameter
$-1<w<-1/3$ are limits for inflation and k-dynamics due to spatial
curvature. Inflation occurs as $w~\rightarrow~-1$ as $\Omega_{\phi}$
grows larger than $\Omega_{k}$ which vanishes. The total density
$\Omega~=~\Omega_{\phi}~+~\Omega_{m}$ $=~1~+~\Omega_{k}~\simeq~1$
is sufficiently close to unity for $\Omega_{k}~=~\rho_{k0}/\rho_{c}$
$=~3M_{pl}^{2}/8\pi(a_{0}H_{0})^{2}~<<~1$. The $k~=~1$ FLRW for
closed spacetime such that $\Omega_{k}$, no matter how small, saturates
the HP at the turn about in the cosmic evolution.

\section{The three phases}

The earliest phase is the pre-inflationary phase given by the FLRW
equation 

\[
\left(\frac{\dot{a}}{a}\right)^{2}=H^{2}~=~\frac{8\pi G}{3M_{p}^{2}}\left(\frac{a_{0}}{a}\right)^{4}~-~\frac{1}{a^{2}},
\]
for a spherical space containing relativistic particles. The spatial
dynamics of the space is $a(\chi_{h})~=~Asin(\chi_{h})$ for $A~=~\sqrt{8\pi Ga_{0}^{4}/3M_{p}^{2}}$.
This solution for $\chi_{h}~\rightarrow~\pi$ where $S/A~\rightarrow~\infty$,
where the singularity at $\chi_{h}~=~0$ is precluded by the Planck
scale cut-off in $a_{0}$. 

During this phase $\Omega_{k}>>\{\Omega{}_{m},~\Omega_{\phi}\}$ and
$\Omega_{k}~\lesssim~1$ the scale factor dynamics is

\[
\dot{a}~\simeq~a_{0}H_{0}\sqrt{\Omega_{k}}
\]
with

\[
\chi_{h}~\simeq~\frac{1}{a_{0}H_{0}\sqrt{\Omega_{k}}}ln\left(\frac{t}{t_{0}}\right),
\]
for $t_{0}~\sim~t_{p}$. Assume the cutoff distance is the string
length $a_{0}~\simeq~\ell_{s}$ and the Hubble inverse time $H_{0}~\simeq~1/\ell_{p}$,
for $\ell_{p}$ the Planck length. Thus $a_{0}H_{0}~\sim~g^{-1/4}$,
for $g$ the string coupling constant. The condition $\chi_{h}~=~\pi/2$
for maximum expansion corresponds to the time 

\[
t~=~t_{0}exp(\chi_{h}a_{0}H_{0}\sqrt{\Omega_{k}})~=~t_{0}exp(\pi/2g^{1/4})
\]
The string parameter $g^{1/4}~\simeq~.06$ computes $t~=~2.3\times10^{11}t_{p}$,
or $t~\simeq~2\times10^{-32}sec$, which agrees with current models.
The $S/A$ ratio is 

\[
\frac{S}{A}\Big|_{\chi_{h}=\pi/2}~=~\sigma\frac{\pi}{4a_{max}^{2}},
\]
for $\sigma$ the constant entropy density. The entropy is given directly
by entropy density $S_{m}~=~\pi\sigma$. The derivative of this function
at $\chi_{h}~=~\pi/2$ is 

\[
\frac{d}{d\chi_{h}}\frac{S}{A}\Big|_{\pi/2}~=~\frac{\sigma}{a_{max}^{2}}~=~\frac{4}{\pi}\frac{S}{A}\Big|_{\pi/2},
\]
giving the entropy bound at the maximal expansion $S~=~kA/4\ell_{p}^{2}$,
the saturation point, where continued evolution surpasses the Bekenstein
bound and $S/A$ diverges. 

The cosmology at $t~\sim~10^{-35}sec$ gives a Bekenstein bound of
$\sim~10^{16}$ bits of information, or $S~\sim~10^{-7}J/K$. The
Boltzmann factor $E/kT~\sim~10^{16}$ equals the quantum phase $E\tau/\hbar$,
for $\tau$ Euclidean time. For the temperature approximately $10^{-8}$
times the Hagedorn temperature $\tau~=~\hbar/kT$ $\sim~10^{-34}sec$,
which corresponds to the onset of inflation and a phase transition.
At this point the energy density of the inflaton field became larger
than the curvature; pre-inflation $\rho_{k}~>>~\rho_{\phi}$ and inflationary
$\rho_{\phi}~>>~\rho_{k}\sim\rightarrow\sim0$. The strict $\Omega_{k}~=~0$
preserves the HP, and the universe converts from $\mathbb{S}^{3}$
to $\mathbb{R}^{3}$. The geometry 

\[
ds^{2}~=~-dt^{2}~+~a^{2}(t)d\Omega_{3}^{2},
\]
transitions to one of greater symmetry. The scale factor $a(t)~=~sin(t/t_{0}$)
is replaced with $a(t)~=~cosh(t\Lambda/3)$ for a de Sitter spacetime. 

The HP protection is a quantum critical phase transition. The FLRW
equation 
\[
\left(\frac{\dot{a}}{a}\right)^{2}=H^{2}~=~\frac{8\pi G}{3}\rho~-~\frac{1}{a^{2}},
\]
for a closed spherical spacetime predicts a large $H$ for $a\ge0$.
The gravity action $S=\int d^{4}x\sqrt{-g}R$ is 
\[
S~=~\int d^{4}x\sqrt{g}\left(\frac{\partial g^{ij}}{\partial t}{\cal G}_{ijkl}\frac{\partial g^{kl}}{\partial t}~-~\frac{\delta W}{\delta g_{ij}}{\cal G}_{ijkl}\frac{\delta W}{\delta g_{kl}}\right)
\]
with the superspace metric,

\[
{\cal {\cal G}}_{ijkl}~=~\frac{1}{2}g^{-1/2}(g_{ik}g_{jl}~+~g_{il}g_{jk}~-~g_{ij}g_{kl}).
\]
The Ricci flow for the manifold is 
\[
\frac{\partial g^{ij}}{\partial t}~=~2R^{ij}~+~\nabla^{i}N^{j}~+~\nabla^{j}N^{i},
\]
and $\partial W/\partial g^{ij}$ constructs a potential term. The
metric for the FLRW gives the relevant superspace metric components

\[
{\cal G}_{iiii}~=~-{\cal G}_{iijj}~=~\frac{1}{2}a.
\]
The metric time derivative is $\partial g^{ii}/\partial t~=~\partial a^{-2}/\partial t$
$=~-2(\dot{a}/a^{3})$ $=~2R^{ii}$ and the kinetic energy is $K~=~-6(\dot{a}/a){}^{2}a^{-3}$.
The quadratic term $(\delta W(g_{ij})/\delta g_{ij})^{2}$ $=~4(1~-~a^{2}/a_{0}^{2})$
gives the potential 
\[
\frac{\delta W}{\delta g^{ij}}{\cal G}{}^{ijkl}\frac{\delta W}{\delta g^{kl}}~=~-6a^{-1}\left(1~-~\frac{a^{2}}{a_{0}^{2}}\right).
\]
The kinetic and potential energy give the Hamiltonian operator, and
the irrelevant factor of 1/6 is ignored. The Hamiltonian acts on the
wave functional $\Psi[g]$ as

\[
\left(\frac{\dot{a}}{a}\right)^{2}~+~a^{2}\left(1~-~\frac{a^{2}}{a_{0}^{2}}\right)~\rightarrow~\left[\hat{\pi}_{a}^{2}~+~a^{2}\left(1~-~\frac{a^{2}}{a_{0}^{2}}\right)\right]\Psi[a]~=~0,
\]
for $\dot{a}/a~\rightarrow~$$\hat{\pi}_{a}~=~-i\partial_{a}$, which
gives this Wheeler-Dewitt equation. The numerical solution for $\Psi[a]$
for $k~=~1$ and $\Omega_{k}~\lessapprox~\Omega$ appears in figure
1. The inflationary phase resets the potential term with constant
$\rho$, which defines a de Sitter space with a greater symmetry and
where $\mathbb{S}^{3}~\rightarrow~\mathbb{R}^{3}$.

\includegraphics[bb=0bp 0bp 650bp 390bp,width=12cm,height=7.5cm]{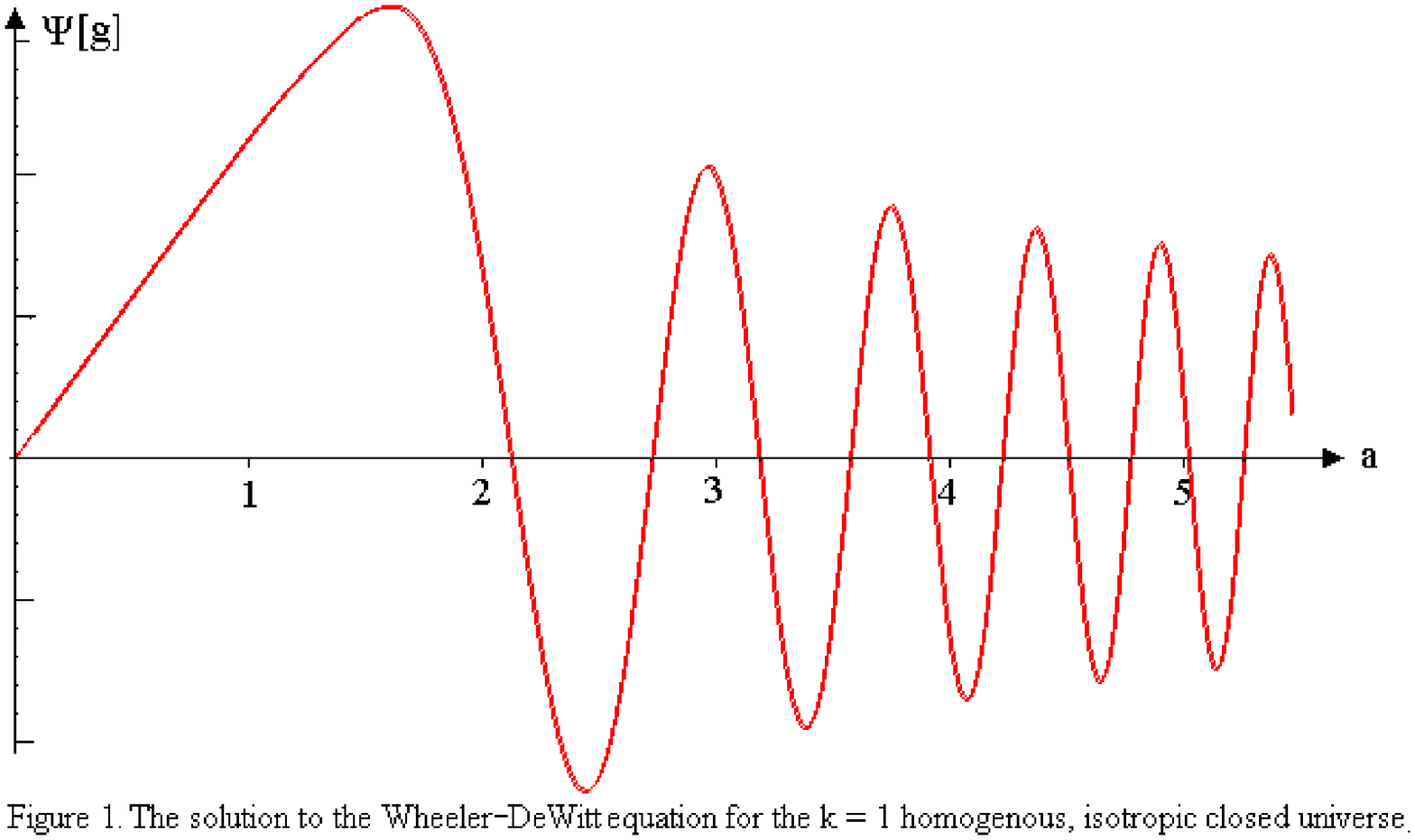}

The transition from {}``k-dynamics,'' with $w~=~-1/3$ with equation
of state $p~=~w\rho$, to the inflaton field, with $w~=~-1$ means
the scale factor is replaced with $a(t)~=~cosh(Ht)/H$. The inflationary
metric expands as $\sim~e^{2\phi}g_{ij}$ with the inflaton field
$\phi$. The Ricci flow $\partial g_{ij}/\partial t~=~2\dot{\phi}g_{ij}$
gives the dynamical equation $\ddot{a}/a~=~4\pi G(\rho~+~3p)$. The
potential function, determined by $W$ above is a polynomial form,
typically $\frac{1}{2}m\phi^{2}$. The potential function for the
evolution with a slowly varying $\phi$ $\partial_{\phi}V(\phi)~\simeq~H^{2}=~\Lambda/3$,
is approximately constant $8\pi G\rho$. The {}``friction term''
$3H\dot{\phi}$ due to dilution the scalar field with the wave equation
in equation 4, so $\dot{\phi}~<~0$. A slowly varying potential defines
a de Sitter spacetime in the inflationary phase of the system.

The next phase of the system is reheating, where $\phi~\simeq~0$.
This is the period of particle creation and the thermalization of
the universe. This is the thermal {}``bang'' in the big bang, after
inflation that lasts $t~\simeq10^{-35}$ seconds, with cosmic expansion
under the influence of relativistic particle or radiation with $w~=~1/3$.
This phase is a bubble nucleation of a pocket universe in the de Sitter
inflating space. The cosmic reheating and particle production governs
inflaton decay 

\[
\ddot{\phi}~+~3H\dot{\phi}~+~\partial_{\phi}V(\phi)~=~-\Gamma\dot{\phi},
\]
for $\Gamma$ the rate of particle production. The dilution of radiation
matter in the universe eventually pushes the universe into a $w=-1$
equation of state after the radiation and matter dominated periods.

The three phases are from a Lifshitz tricritical quantum critical
point, such as seen in condensed matter physics {[}6{]}. The first
corresponds to a dynamical $\phi$, which is the pre-inflationary
phase. The next is the high potential phase or $\rho(\phi)\sim10^{72}GeV^{4}$
and the last phase with small density $\rho(\phi)\sim10^{-48}GeV^{4}$,
corresponding to the current cosmological constant. This Lifshitz
triple point is illustrated below, where the ordinate and abscissa
are energy (temperature) and pressure. The system evolves from the
top left as the closed cosmology in the pre-inflationary phase, to
the phase corresponding to inflation, and the potential collapses
in the bottom right which produces particles in the reheating phase.
The scale of the diagram and the path drawn are heuristics.

\includegraphics[bb=0bp 0bp 510bp 350bp,width=9cm,height=6cm]{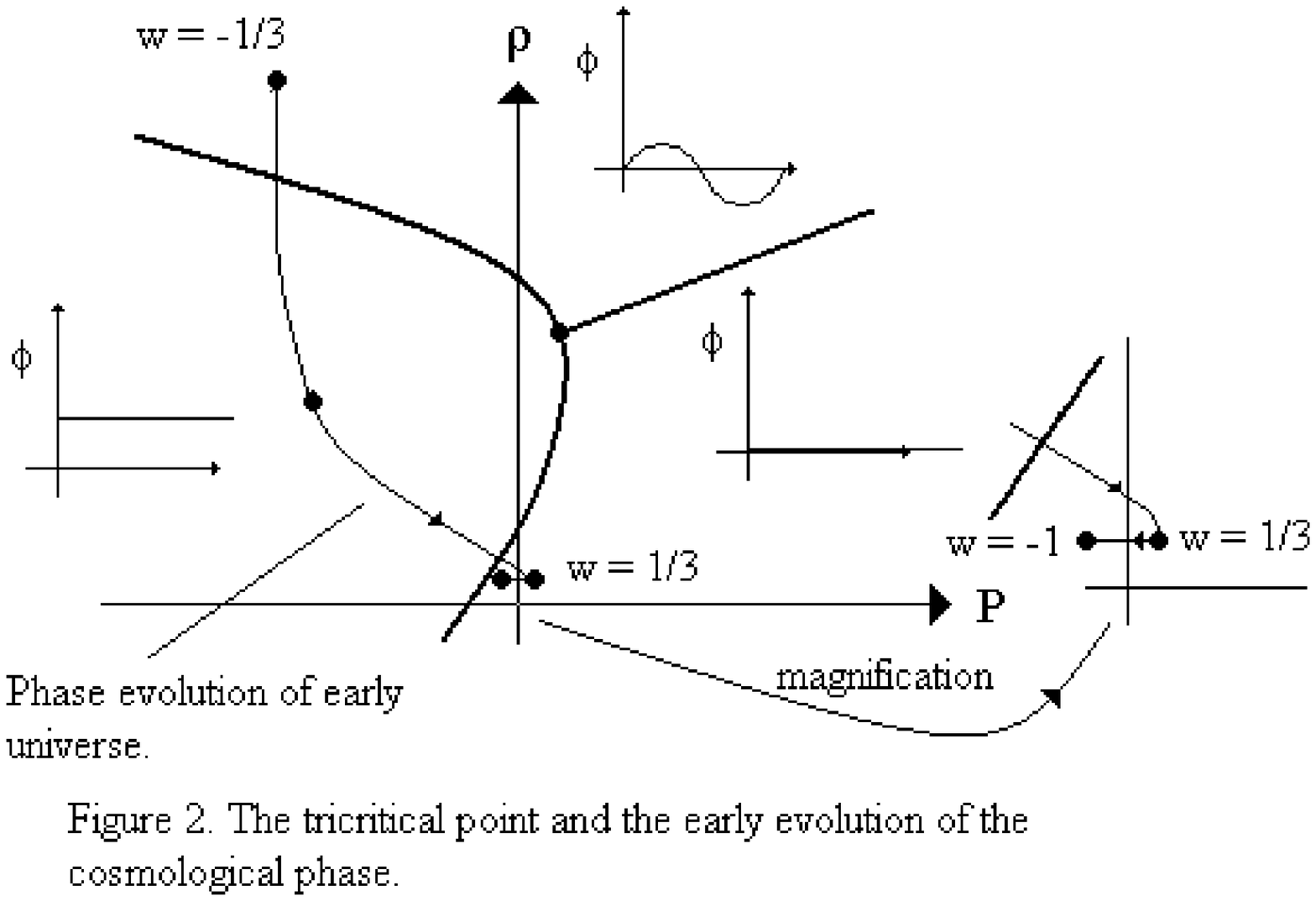}

\section{Quantum tricritical phases}

The HP violation is avoided by a quantum critical point and a phase
transition to the inflationary phase, which enters then into another
phase, which is the reheating of the universe and the current cosmic
state. The Lifshitz tricritical point connects the scalar field $\phi$
in a spatially modulated phase with the phases $\phi~=~const$ and
$\phi~\gtrsim~0$. The spatially modulated phase is the scalar or
inflaton field in the closed spacetime configuration. The next phase
is a large vacuum energy with $\phi~\simeq~const$, and inflation.
The final transition to $\phi~\gtrsim~0$ reheating phase is bubble
nucleation of pocket universes. The tricritical point in connection
to relativity is explored by Horava{[}7{]}, with the reparameterizations
$t~\rightarrow~b^{z}t$ and $x~\rightarrow~bx$, for $z$ the critical
exponent, which give conformal or Ricci flow $g_{ij}~\rightarrow$
$e^{2\phi}g_{ij}$, for $z~=~1$.

The ground state wave functional $\Psi[\phi(x)]~=~exp(-W/2)$ has
dynamics given by 

\[
S~=~\frac{1}{2}\int dtd^{n}x\Big(\dot{\phi^{2}}~-~(\frac{1}{2}\Delta\phi)^{2}\Big).
\]
The functional $W$ for k-dynamics gives the critical exponent $x~\rightarrow~b^{[\phi]}x$
is then $[\phi]~=~0$. This is the spatially modulated phase determined
by $W$. The inflationary phase with a changed critical dimension
and conformal flow $x_{i}~\rightarrow~x_{i}e^{\phi}$. The critical
exponent $x_{i}~\rightarrow~x_{i}b^{[x]}$ defines the factor $b~=~e^{2\phi}$
scale factor with scale weight $[x]~=~1$. The conversion of the closed
sphere $\mathbb{S}^{3}$ to $\mathbb{R}^{3}$ in a simple model is
a stereographic projection. Cartesian coordinates of $\mathbb{S}^{3}$,
such at$(x_{1},x_{2},x_{3},z)$ for $z$ some fictional embedding
spatial coordinate, map as $x_{i}~\rightarrow~x_{i}/(1~-~z)$. We
express the coordinate $z$ according to a field $\phi$ as $1/(1~-~z)~=~e^{\phi}$
so conformal dynamics is a manifestation of topological change $\mathbb{S}^{3}~\rightarrow~\mathbb{R}^{3}$.

The renormalization group flow follows from a conformal rescaling,
which has a rich structure stemming from Zamolodchikov {[}8{]} and
the Cardy a-conjecture {[}9{]}. The Ricci flow $\partial g_{ij}/\partial t=2\dot{\phi}g_{ij}$
computes the kinetic energy term 
\[
\frac{\partial g_{ij}}{\partial t}{\cal G}^{ijkl}\frac{\partial g_{kl}}{\partial t}~=~4\left(\frac{\partial\phi}{\partial t}\right)^{2}g_{ij}g_{kl}{\cal G}^{ijkl}~=~\frac{k}{2}\left(\frac{\partial\text{\ensuremath{\phi}}}{\partial t}\right)^{2}
\]
 for $k=2g_{ij}g_{kl}{\cal G}{}^{ijkl}$ and $\phi$ rescaled to $\phi/2$.
The potential energy term is 
\[
\frac{\delta W}{\delta g^{ij}}~=~\frac{\delta W}{\delta\text{\ensuremath{\phi}}}\frac{\partial\phi}{\partial g^{ij}}~=~\frac{1}{2}\frac{\delta W}{\delta\phi}g_{ij}
\]
 so 
\[
\frac{\delta W}{\delta g^{ij}}G^{ijkl}\frac{\delta W}{\delta g^{kl}}~=~\frac{1}{4}\left(\frac{\delta W}{\delta\text{\ensuremath{\phi}}}\right)^{2}g_{ij}g_{kl}G^{ijkl}~=~\frac{k}{2}\left(\frac{\delta W}{\delta\phi}\right)^{2}.
\]
The Lagrangian 
\[
{\cal L}~=~\frac{k}{2}\left(\frac{\partial\phi}{\partial t}\right)^{2}~-~\frac{k}{2}\left(\frac{\delta W}{\delta\text{\ensuremath{\phi}}}\right)^{2}.
\]
with the functional $W=W[\phi]$ of the form $\dot{\phi^{2}}~-~(\frac{1}{2}\Delta\phi)^{2}$
for $\Delta\phi~=~P(\phi)$, a polynomial so 

\[
{\cal L}~=~\frac{k}{2}\left(\frac{\partial\phi}{\partial t}\right)^{2}\text{~-~}k\sum_{n}\frac{a_{n}}{n}\phi^{n}.
\]
A possible first choice of a potential is 
\[
\left(\frac{\delta W}{\delta\text{\ensuremath{\phi}}}\right)^{2}~=~V(\phi)~=~\frac{a_{2}}{2}\phi^{2}~+~\frac{a_{4}}{4}\phi{}^{4}
\]
which is zero for $\phi=0$ and $\phi=\sqrt{-2a_{2}/a_{4}}$ for $a_{2}<0$.
The Thomas-Fermi approximation {[}10{]} {[}11{]} for phase transitions
employs a cubic term 
\[
V(\phi)~=~\frac{a_{2}}{2}\phi^{2}~-~\frac{a_{3}}{3}\phi^{3}~+~\frac{a_{4}}{4}\phi^{4}
\]
 For $a_{2}<a_{3}^{2}/4a_{4}$ there is a third minimum 
\[
\phi_{0}~=~\frac{a_{3}}{2a_{4}}\Big(1~+~\sqrt{1~-~4a_{2}a_{4}/a_{3}^{2}}\Big)
\]
 The potential for different values of $a_{3}$ appears in figure
3

\includegraphics[bb=0bp 0bp 378bp 250bp,width=8cm,height=5cm]{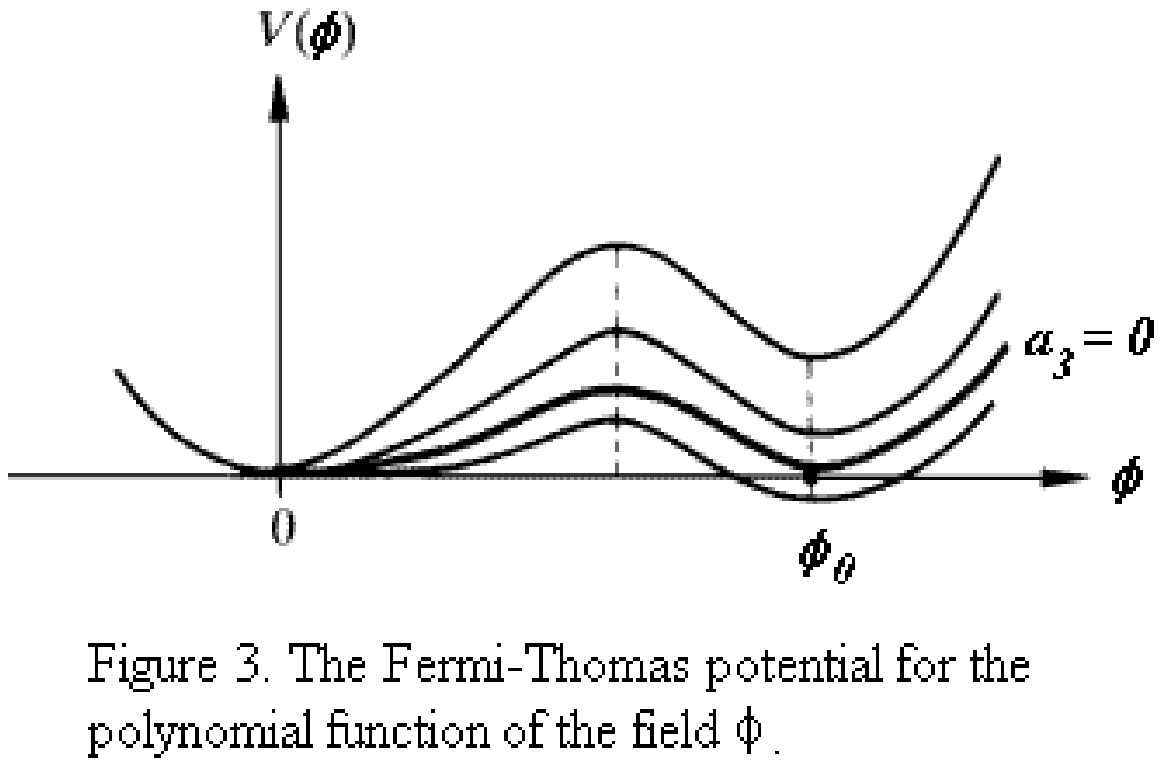}

The field exhibits a phase transition jump to $\phi~=~0$ from $\phi_{0}~=~2a_{3}/3a_{4}$.
The coherence length of the $\phi$-fluctuations for $a_{3}=0$ is
computed from the nonzero expectation of the field $\langle\phi\rangle~=~\phi_{0}~=~\sqrt{-a_{2}/a_{4}}$,
which corresponds to a Higgs-like mass, and the coherence length $\xi~=~1/\sqrt{-2a_{2}}${[}12{]}.
In superconductivity this connects the Meissner-Higgs mass term to
the penetration depth of a magnetic field $\lambda~\sim~1/\phi$.
The ratio of the two length scales is the Ginsburg parameter $k~=~\lambda/\sqrt{2}\xi$,
which is large for type I superconductors and small for type II. Including
the cubic term the coherence length is 
\[
\xi~=~\frac{1}{\sqrt{a_{2}~+~3a_{4}\phi^{2}\text{~-~}2a_{3}\phi}},
\]
which around $\phi~=~0$ is $\xi~=~1/\sqrt{a_{2}}$. For $a_{3}~=~0$
the field jumps to $\phi_{0}~=~\sqrt{2a_{3}/3a_{4}}$ and the $\phi$-fluctuation
length jumps to 
\[
\xi_{1}~=~\frac{3}{a_{3}}\sqrt{\frac{a_{4}}{2}}
\]
 corresponding to a first order phase transition.

The expected field $\phi_{0}~=~\langle\phi\rangle$$~=~\sqrt{2a_{3}/3a_{4}}$
is the critical parameter for inflation and HP protection. The metric
at this critical value $g_{ij}~=~exp(2\phi_{0})~\sim~a(t^{\prime})$
cuts off at the value of $\pi/2~<~\theta~<~\pi$ where the entropy
bound is saturated. The phase change initiates inflation or de Sitter
dynamics with a large vacuum energy density and pressure. 

Transition to the third phase with $\phi~\gtrsim~0$ causes production
of particles. The vacuum energy becomes very small, with inflationary
expansion with a larger e-fold time $t_{e}~=~1/H$. This is bubble
nucleation {[}13{]} transpires on the de Sitter spacetime inflating
on a time scale $t~\simeq~\ell_{s}$, where each bubble subsequently
expands on a time scale $1/H$. This is the first multiverse scenario
where the space $\mathbb{R}^{3}$ rapidly inflates and local regions
tunnel into a lower energy configuration. These regions define pocket
universes within the Guth-Linde-Vilenkin multiverse scenario {[}14{]}{[}15{]}.

\section{Final Remarks}

This potentially connects \char`\"{}braney\char`\"{} physics with
the standard cosmological model. A cosmology is then the stringy content
of a D3-brane. Some models have cosmologies generated by the collision
between two branes {[}16{]}. However, it may be quite the opposite.
It is possible that the elementary particles in our universe are truly
point-like according to zero electric dipole moments, as experiments
are beginning to suggest, with dipole moment $0.07\pm0.07\times10^{\lyxmathsym{\textminus}26}e\cdot cm${[}17{]},
which is close to some SUSY predictions of an electric dipole. The
strings for these particles connect two D3-branes. We might think
of some field flux which causes open strings to connect two branes
together, instead of both endpoints on a single brane. The branes
may be dynamical and are moving apart, which stretches the string.
The string may break, but rather than having free endpoints they are
attached to a new braney object. This object has the topology of a
sphere, which transforms its topology to $\mathbb{R}^{3}$. Consider
a foliation of D3-branes in expanding bulk space which pulls branes
apart. As teh brane foliation separates gaps are filled by the generation
of new branes, which contain nascent inflationary cosmologies, which
in turn generate pocket universes. The tricritical phase transitions
are associated with both the generation of the brane and subsequent
bubble nucleation on it.

\includegraphics[bb=0bp 0bp 520bp 260bp,width=10cm,height=5cm]{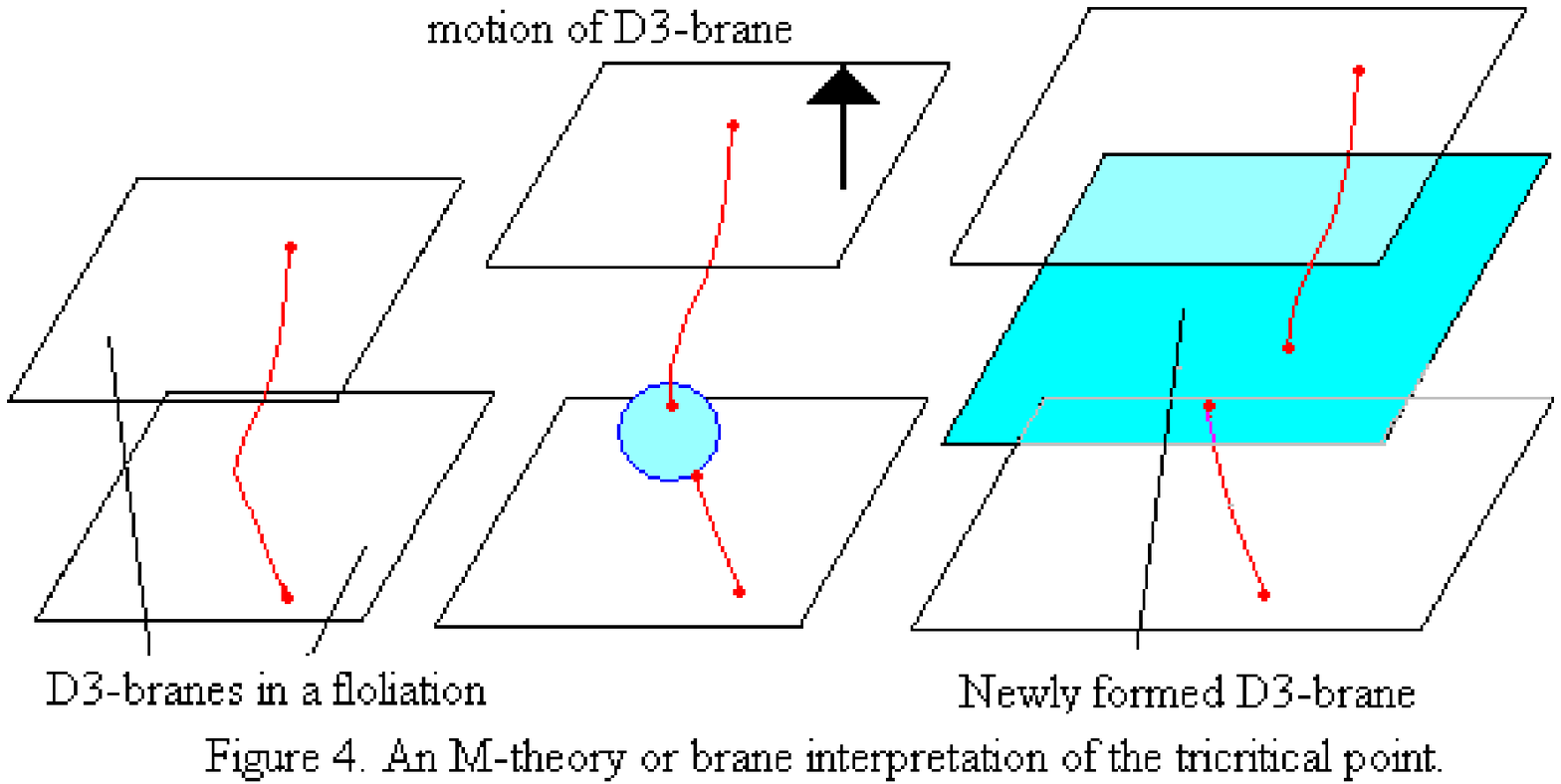}

Dp-branes have a large $N\hbar$ content and are classical. The emergent
small $\mathbb{S}^{3}$ with the end points of the cut string has
more quantum mechanical content. $\mathbb{S}^{3}~\rightarrow~\mathbb{R}^{3}$
is a transition to classical mechanics for the large D3-brane. The
tricritical phase transition from a varying field to constant field
transforms $\mathbb{S}^{3}$ into a classical spacetime $\mathbb{R}^{\ensuremath{3}}$.
This middle phase is the inflationary period of the universe. The
final phase is bubble nucleation and the formation of pocket universes.
This reheating phase produces the radiation, matter and small accelerated
phases of the universe we observe. 

The reheating phase of the universe breaks the Weyl transformation
or conformal scaling of the inflationary period. The large inflationary
$\Lambda$ is reduced to near zero, which accompanies the breaking
of conformal symmetry if $\Lambda~=~0$ identically. The dilaton in
this setting emerges in vacua with spontaneous broken conformal symmetry.
However, physically it is not identically zero. It is possible high
energy scale physics of inflation is dual to the low energy physics
of the cosmological constant of the current universe in a T-duality
or $k~\rightarrow~1/k$. The high energy physics is $k~\sim~M_{pl}$,
and the low energy physics is $k~\rightarrow~K^{2}/k~\sim~$ $M_{\Lambda}~=~10^{-12}GeV$.
The implicit constant at work here is the intermediary momentum or
mass $K~=~\sqrt{M_{\Lambda}M_{pl}}$ $\sim~10^{3}GeV$, or about the
LHC energy scale. The dual when written according to proper units
is $M_{pl}~\rightarrow~K^{2}/M_{pl}$ $\sim~M_{\Lambda}$. This is
the value of the vacuum energy for the universe and the cosmological
constant. This suggests some underlying UV/IR duality to renormalization
group flow.

\section{References}

{[}1{]} W. Fischler and L. Susskind, and R. Bousso, $Rev.~Mod.~Phys.$
${\bf 74}$, 825 (2002).hep-th/9806039.

{[}2{]} J.D. Bekenstein, $Phys.~Rev.$ ${\bf D23}$, 287 (1981).

{[}3{]} M. Li, $Phys.~Lett.$ ${\bf B603}$, 1 (2004).

{[}4{]} E. J. Copeland, M. Sami, S. Tsujikawa, $Int.~J.~Mod.~Phys.$
${\bf D15}$, 1753 (2006)

{[}5{]} V. H. Cardenas, \char`\"{}Inflation as a response to protect
the Holographic Principle,'' arXiv:0908.0287v1 {[}gr-qc{]}

{[}6{]} T. Misawa, Y. Yamaji, M. Imada, \char`\"{}YbRh2Si2: Quantum
tricritical behavior in itinerant electron systems,\char`\"{} $J.~Phys.~Soc~Jpn$.
${\bf {\bf 77}}$ (2008) arXiv:0710.3260v2 {[}cond-mat.str-el{]}

{[}7{]} P. Horava, \char`\"{}Quantum Gravity at a Lifshitz Point,\char`\"{}
$Phys.~Rev.$ ${\bf D79}$ (2009) arXiv:0901.3775v2 {[}hep-th{]} 

{[}8{]} A.B Zamolodchikov, \char`\"{}Irreversibility' Of The Flux
Of The Renormalization Group In A 2-D Field Theory,\char`\"{} $JETP~Lett.$
${\bf 43}$ 730\textendash{}732 (1986)

{[}9{]} G. Feverati, \char`\"{}Exact (d)-\textgreater{}(+)\&(-) boundary
flow in the tricritical Ising model,\char`\"{} $J.~Stat.~Mech.$ 0403:P03001,(2004)
arXiv:hep-th/0312201v2

{[}10{]} E. Fermi, \char`\"{}Un Metodo Statistico per la Determinazione
di alcune Prioprietà dell'Atomo,\char`\"{} $Rend.~Accad.~Naz.~Lincei$
${\bf 6}$, 602\textendash{}607 (1927) http://babel.hathitrust.org/cgi/pt?

seq=339\&view=image\&size=100\&id=mdp.39015001321200\&u=1\&num=278. 

{[}11{]} L. H. Thomas, \char`\"{}The calculation of atomic fields,\char`\"{}
$Proc.~Cambridge~Phil.~Soc.$ ${\bf 23}$ 5 542\textendash{}548.(1927) 

{[}12{]} C. Kittel, $Introduction~to~Solid~State~Physics$, John Wiley
\& Sons. pp. 273\textendash{}278 (2004)

{[}13{]} S. Coleman, F. De Luccia, \char`\"{}Gravitational effects
on and of vacuum decay,\char`\"{} $Phys.~Rev.$ ${\bf D21}$: 3305
(1980).

{[}14{]} A. D. Linde, \char`\"{}Eternally Existing Self-Reproducing
Chaotic Inflationary Universe \&,\char`\"{} $Physics~Letters$ ${\bf B175}$
4: 395\textendash{}400 (1986)

{[}15{]} A. Borde, A. H. Guth, A. Vilenkin, \char`\"{}Inflationary
spacetimes are not past-complete,\char`\"{} $Phys.~Rev.~Lett.$ ${\bf 90}$
(2003)

{[}16{]} J. Khoury, B. A. Ovrut, P. J. Steinhardt and N. Turok, \char`\"{}The
ekpyrotic universe: Colliding branes and the origin of the hot big
bang\char`\"{}, $Phys.~Rev.$ ${\bf D64}$, 123522 (2001) arXiv:hep-th/0103239.

{[}17{]} J. J. Hudson, D. M. Kara, I. J. Smallman, B. E. Sauer, M.
R. Tarbutt, E. A. Hinds, \char`\"{}Improved measurement of the shape
of the electron\char`\"{}. $Nature$ ${\bf 473}$, 7348 493\textendash{}496
(2001)
\end{document}